# Very strong intrinsic supercurrent carrying ability and vortex avalanches in (Ba,K)Fe$_2$As$_2$ superconducting single crystals

Xiao-Lin Wang[1*], S. R. Ghorbani[1,2], Sung-Ik Lee[3**], S.X. Dou[1], C. T. Lin[4], T. H. Johansen[5], Z. X. Cheng[1], G. Peleckis[1], K. Muller[6], M. Shabazi[1], G.L. Sun[4], and D.L. Sun[4]

[1]Institute for Superconducting and Electronic Materials, Faculty of Engineering, University of Wollongong, Wollongong, New South Wales 2522, Australia
[2]Department of Physics, Tarbiat Moallem University of Sabzevar, P.O. Box 397, Sabzevar, Iran
[3]National Creative Research Initiative Center for Superconductivity, Department of Physics, Sogang University, Seoul 121-742, Republic of Korea
[4]Max Planck Institute for Solid State Research, Heisenbergstr 1, 70569 Stuttgart, Germany.
[5]Department of Physics, University of Oslo, POB 1048 Blindern, 0316 Oslo, Norway
[6]CSIRO, CMSE, Lindfield, NSW 2070, Australia

**A high critical current density, $J_c$, upper critical field, $H_{c2}$, and irreversibility field, $H_{irr}$, a high superconducting transition temperature, $T_c$, strong magnetic flux pinning, good grain connectivity, and isotropic superconductivity are the major physical requirements for superconducting materials used in practical applications operating at low, and in particular, high magnetic fields. The conventional low $T_c$ superconductors, where $H_{c2}$ is also small, can only carry large $J_c$ at very low temperatures. The cuprate high $T_c$ superconductors suffer from poor grain connectivity and easy melting of the vortex lattice, leading to small $J_c$ in high magnetic fields at relatively high temperatures. For MgB$_2$ superconductor with $T_c$ of 39 K, $H_{irr}$ is far below $H_{c2}$, and $J_c$ drops quickly with both field and temperature, preventing its use above 20 K. The newly discovered Fe-based superconductors[1-7] show $T_c$ as high as 55 K and $H_{c2}$ above 200 T, in combination with a small anisotropy for REFeAsO$_{1-x}$F$_x$ (RE-1111 phase, with RE a rare earth element)[8] and an almost isotropic superconductivity for (Ba,K)Fe$_2$As$_2$ (122 phase).[9] These properties make the Fe-based superconductors extremely promising candidates for high magnetic field applications at relatively high temperatures. The current carrying ability of these superconductors at high fields and temperatures is largely determined by the flux pinning strength and the behavior of the vortex matter. Therefore, the determination of their intrinsic vortex pinning strength is a central issue from both an applied and a fundamental perspective. The present investigation shows that single crystals of (Ba,K)Fe$_2$As$_2$ with $T_c$ = 32 K have a pinning potential, $U_0$, as high as $10^4$ K, with $U_0$ showing very little field dependence. In addition, the (Ba,K)Fe$_2$As$_2$ single crystals become isotropic at low temperatures and high magnetic fields, resulting in a very rigid vortex lattice, even in fields very close to $H_{c2}$. The rigid vortices in the two dimensional (Ba,K)Fe$_2$As$_2$ distinguish this compound from 2D high $T_c$ cuprate superconductors with 2D vortices. The very strong intrinsic supercurrent carrying ability and flux pinning observed in the (Ba,K)Fe$_2$As$_2$ superconducting single crystals make this compound very promising for future applications in high magnetic fields.**

Both 1111 and 122 phase compounds have typical two dimensional (2D) crystal structures. In RE-1111 phase, where RE is a rare earth element, the FeAs superconducting layers areseparated by insulating LaO layers,[10] while in Ba(K)-122 phase, the FeAs layer is sandwiched between conductive Ba layers.[5] It is expected that the 122 phase containing two FeAs layers would have small anisotropy and thus higher intrinsic pinning compared to the single layer 1111 phase. Co doped BaFe$_2$As$_2$ single crystal shows an anisotropy of 1-3 and upper critical field values of $H_{c2}(H//ab)$ = 20 T and $H_{c2}(H//c)$ = 10 T at 20 K, with $dH_{c2}/dT \approx 5$ T/K.[11] For single crystals of the optimally doped Ba(Fe$_{1-x}$Co$_x$)$_2$As$_2$ with x = 0.074 and critical temperature $T_c \approx 23$ K, the anisotropy of the upper critical field, $\gamma = H_{c2}^{ab}/H_{c2}^{c}$, is in the range of 2.1 to 2.6, and the critical current density, $J_c$, is over $10^5$ A/cm$^2$ and $3\times10^5$ A/cm$^2$ at 5 K for H//ab and H//c, respectively.[11] The $H_{c2}$ of (Ba$_{0.55}$K$_{0.45}$)Fe$_2$As$_2$ measured under pulsed magnetic fields exceeds 60 Tesla at 14 K.[9] The anisotropy of $H_{c2}$ is moderate (~ 3.5 close to $T_c$), and it drops with decreasing temperature, becoming isotropic at low temperatures[9]. Underdoped BaFe$_2$As$_2$ with $T_c$ of 25 K shows an anisotropy of 3-4.[12]

Melting of the vortex lattice is the major cause for the poor $J_c$ in high field and high temperature for both MgB$_2$ and cuprate superconductors. Study of the intrinsic flux pinning and phase diagram of the Fe-based superconductors is crucially important for understanding their vortex matter and their capability for carrying supercurrent, which will determine whether or not these new Fe-superconductors can be useful for practical applications, even though they have already been shown to have very high $H_{c2}$ in both 1111 and 122 compounds. Another important issue for practical applications is the magnetic flux jumping, which reflects the vortex stability and is an indication of high $J_c$. Its investigation is closely relevant to understanding the complexity of the vortex matter in the mixed phase of superconductors. Flux jumping has been observed in thin film forms of some type II superconductors, such as Nb$_3$Sn,[13] and MgB$_2$.[14] However, flux jumping has not yet been reported in any Fe-based superconductors so far.

We report that the vortices in optimally doped (Ba,K)Fe$_2$As$_2$ crystal are extremely rigid, with giant pinning potentials that are as high as $10^4$ K and almost field independent. The vortices are pinned nearly three dimensionally at low temperatures and high magnetic fields. The rigid vortices distinguish the two dimensional (Ba,K)Fe$_2$As$_2$ from 2D high $T_c$ cuprate superconductors with 2D vortices. Flux jumping due to high $J_c$ was also observed in large samples and very low temperatures..

The 122 crystals used in the present work was grown using a flux method. High purity elemental Ba, K, Fe, As, and Sn were mixed in a mol ratio of (Ba$_{1-x}$K$_x$Fe$_2$As$_2$):Sn = 1:45-50 for the self flux. A crucible with a lid was used to mini-mize the evaporation loss of K as well as that of As during growth. The crucible was sealed in a quartz ampoule filled with Ar and loaded into a box furnace. The details of the crystal growth are given in Ref. 15.

The temperature dependence of the resistance of a $Ba_{0.72}K_{0.28}Fe_2As_2$ crystal for field applied parallel to the *ab* plane and to the *c*-axis is shown in Fig. 1. The resistance drops to zero at $T = 33$ K in zero magnetic field, indicating that the crystal is close to optimum doping. It can be seen that the onset $T_c$ drops very slowly with increasing magnetic field, and the $T_c(R=0)$ shows the same trend, indicating that the transition width remains almost constant for both H//*ab* and H//*c* when the field increases (see the inset in Fig 1(a)). When the field changes from 0 up to 13 T, the shape of R(T) changes very little. This behavior is reminiscent of the magneto-transport behavior of conventional low-$T_c$ superconductors, but significantly different from that of cuprate high-$T_c$ superconductors. In cuprate high-$T_c$ superconductors, the $T_c$ onset temperature does not change much, but the $T_c(R=0)$ shifts to low temperature very quickly with field. The constant transition width, $\Delta T_c$, is also quite different from what is observed in $NdFeAs_{0.2}OF_{0.8}$ single crystals, which show a broadening of $T_c(R=0)$ as field increases.[8] $H_{c2}$ is defined as the field at which the resistance R = 0.9 $R_n$, where $R_n$ is the normal state resistance, while the irreversibility field $H_{irr}$ is defined by R = 0.1 $R_n$. The $H_{c2}$ and $H_{irr}$ obtained for both H//*ab* and H//*c* are denoted as $H_{c2}^{ab}$, $H_{c2}^c$, $H_{irr}^{ab}$, and $H_{irr}^c$, respectively.

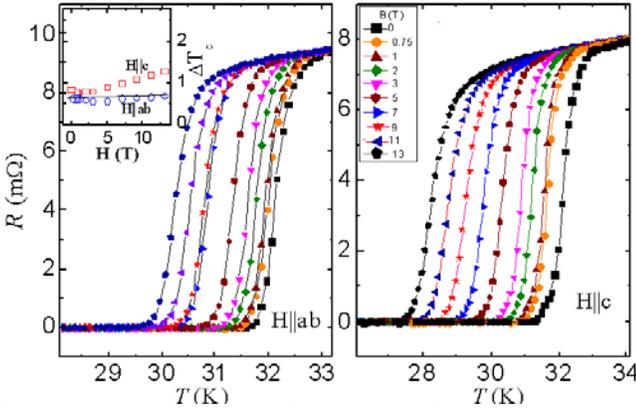

Fig. 1. Temperature dependence of the resistance of a $Ba_{0.72}K_{0.28}Fe_2As_2$ crystal measured for H//*ab* and H//*c* in different magnetic fields, $0 \leq H \leq 13$ T. Inset in the left panel shows the transition width as a function of magnetic field for both directions of applied field.

It can be seen from Fig. 2 that the $H_{c2}$ values for the (BaK)$Fe_2As_2$ are quite high ($H_{c2}^{ab} = 13$ T at 30.7 K and $H_{c2}^c = 13$ T at 29 K). The slope, $|dH_{c2}/dT|$, is 7.5 and 4.4 for the *ab* and the *c* direction, respectively, for our $Ba_{0.72}K_{0.28}Fe_2As_2$ crystal. From Fig. 2, we can see that $H_{irr}$ is very close to $H_{c2}$ in both field directions, indicating that the intrinsic flux pinning is very strong in the (Ba,K)$Ba_2As_2$ crystal.

The $H_{c2}(0)$ was estimated by using the Werthamer-Helfand-Hohenberg (WHH) formula: $H_{c2}(0) = -0.69$ $T_c \cdot (dH_{c2}/dT)$, with $dH_{c2}/dT$ at $T = T_c$. From the slopes, $H_{c2}^{ab}(0)$ and $H_{c2}^c(0)$ of 170 and 100 T were obtained for our (Ba,K)$Fe_2As_2$ crystal. Furthermore, from the Ginzburg-Landau (GL) equation, $H_{c2}(T) = H_{c2}(0) \cdot (1-t^2)/(1+t^2)$, where $t = T/T_c$ is the reduced temperature. It was found that $H_{c2}^{ab}(0)$ and $H_{c2}^c(0)$ is 195 and 110 T, respectively. If we take into account the positive curvature of $H_{c2}$, then $H_{c2}(0)$ of over 200 T is easily achievable. The anisotropy $\gamma$ ($H_{c2}^{ab}/H_{c2}^c$) of the 122 crystal was determined to be 2-3.5 for $T > 30.5$ K, as can be seen in the inset of Fig. 2. It should be noted that $\gamma$ will further decrease with decreasing temperature and will reach 1 at $T = 0$ K, as estimated using $\gamma(T) = \gamma(T=0)(1-T/T_c)^n$, where $n = -0.3$. This trend in $\gamma$ vs. T is opposite to that of Sm-1111[15] and $MgB_2$[16] superconductors, indicating that the 122 crystal shows large and significantly different multiband effects.

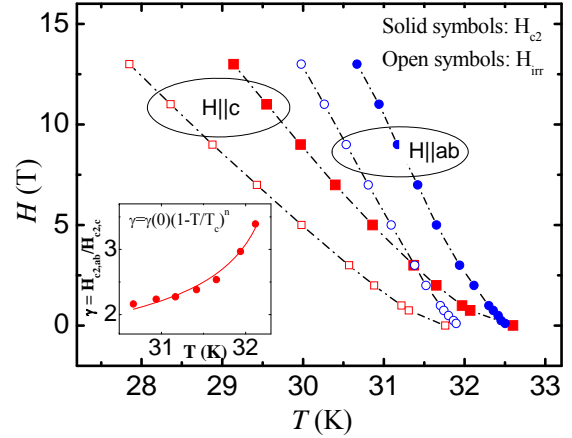

Fig. 2. $H_{c2}$ and $H_{irr}$ for both H//ab and H//c as a function of temperature. Inset: the anisotropy $\gamma$ ($H_{c2}^{ab}/H_{c2}^c$) as a function of temperature.

According to the thermally activated flux flow model, the temperature dependence of the resistivity, R(T), is described by the equation R(T,B) = $R_0$exp[-$U_0/k_BT$], where $R_0$ is a parameter, $k_B$ is the Boltzmann constant, and $U_0$ is the vortex pinning potential, and we assume $U(T,H) = U_0(H)(1-t)$. This pinning potential can be obtained from the slope of the linear part of an Arrhenius plot, log R(T,B) versus 1/T, in a temperature interval below $T_c$. In Fig. 3, we plot the data as log R vs. 1/T. The linearity of log R vs. 1/T in the low temperature region indicates the thermally activated behavior of the resistance. The slope of the curve is the pinning potential $U_0$. The best fit of the experimental data yields values of the pinning potential ranging from $U_0 = 9100$ K and 5900 K for H//*ab* and H//*c*, respectively, at the low field of 0.1 T. These values are much higher than the reported values of $U_0 = 2000-3000$ K for $NdO_{0.82}F_{0.18}FeAs$ single crystals.[8] For comparison, we also include the $U_0$ for Bi-2212[17] and Y-123[18] single crystals in Fig.4, even though their Tc are different.

It has been reported that the pinning potential of bismuth strontium calcium copper oxide (BSCCO) crystals exhibits a power-law dependence on magnetic field, $U_0(B) \propto B^{-n}$, with $n = 1/2$ for B < 5 T and $n = 1/6$ for B > 5 T for H//*c*.[17,19] However, for the (BaK)$Fe_2As_2$ crystal, $U_0$ drops very slowly with field and scales as $B^{-0.09}$ and $B^{-0.13}$ for H//*ab* and H//*c*, respectively (Fig. 4). This means that the $U_0$ is almost field independent, which is a very remarkable result. The values of the field independent $U_0$ for (Ba,K)$Fe_2As_2$ are 3-4 times larger than that of Bi-2212[19] and 10 times greater than that of Bi-2223[17] crystals. These values are also more than three times higher than for $NdO_{0.82}F_{0.18}FeAs$ crystals.[8] The $U_0$ for $(Ba,K)Ba_2As_2$ crystal is even one order of magnitude higher than that of Y-123 crystal for fields above 1 T.[18] The value of $U_0(B)$ obtained from $(Ba,K)Fe_2As_2$ single crystal is a record high compared to that of any other superconductor in single crystal form. Higher pinning potential could be expected in thin films of the $(Ba,K)Fe_2As_2$ compound.

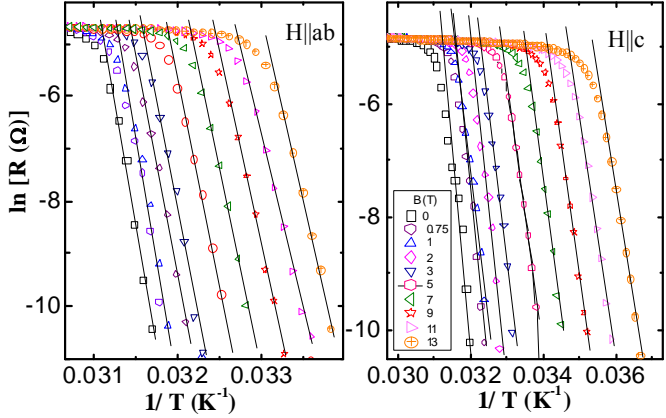

Fig. 3 Log resistance versus 1/T for field parallel (left) and perpendicular (right) to the ab direction.

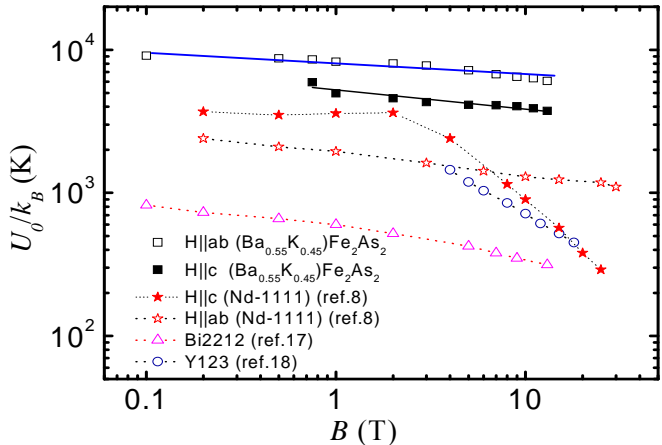

Fig. 4. Field dependence of $U_0(B)$ for $Ba_{0.72}K_{0.28}Fe_2As_2$ crystal $(BaK)Fe_2As_2$.

The magnetic behavior of $(Ba,K)Fe_2As_2$ was investigated using a small sample shaped as a long thin strip, which was obtained by cleaving a single crystal along the *ab*-plane. The left inset in Fig. 5 is a magneto-optical image of the flux penetration in the sample, which was collected while applying a perpendicular field of 50 mT at 20 K. It is evident that the sample is of high uniformity, i.e., without any microcracks or weak links perturbing the flow pattern of the shielding current. Such regular flux patterns allow precise measurements of the critical current density using the Bean critical state model formula for partial penetration in the long thin strip geometry[20]

$$\cosh(\pi H / J_c d) = w/(w - 2a),$$

where *a* is the advancement of the flux front into the strip at an applied field *H*, and *w* and *d* are the width and thickness of the strip, respectively. The obtained values for $J_c$ in the temperature range of 10 to 25 K are plotted in the right inset of Fig. 5.

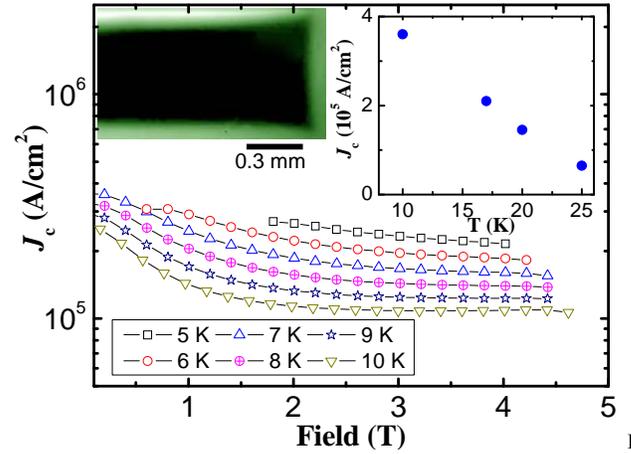

Fig. 5. The $J_c$-field dependence obtained from the M-H loops (Fig. 6) measured on a large $Ba_{0.72}K_{0.28}Fe_2As_2$ crystal. Left inset: magneto-optical image of the flux penetration into the right half of the crystal. Right inset: critical current density versus temperature of a small $(Ba,K)Fe_2As_2$ single crystal shaped as a thin rectangular bar.

From the perspective of practical applications, it is very crucial to investigate if the $(Ba,K)Fe_2As_2$ can carry high supercurrents in high magnetic field and whether or not $J_c$ is stable in larger samples at relatively high temperatures. Magnetic hysteresis loops (Fig. 6) were collected for one of the bigger $(Ba,K)Fe_2As_2$ crystals ($4.2 \times 2.85 \times 0.15$ mm$^3$) in different fields and temperatures down to 5 K. The critical current density was obtained from the vertical width, $\Delta M$, of the magnetization loop using the Bean model interpretation for fully penetrated states,

$$J_c = 20\Delta M / w(1-w/3l),$$

where *l* is the length of the sample, The resulting critical current density versus field is plotted in Fig. 5.

The critical current density is as high as $10^5$ A/cm$^2$. As has been shown for the higher pinning potentials in $(Ba,K)Fe_2As_2$ in the high field limit, we could not see any significant decrease in this value in the field dependence. At 5 K, the $J_c$ value is $2.7 \times 10^5$ A/cm$^2$ at $B = 2$ T, and it only decreases to $2.2 \times 10^5$ A/cm$^2$ in B = 4 T. With increasing temperature, the $J_c$ initially begins to decrease slowly at low field ($B < 2$) and then tends to saturate at high magnetic fields. These results for the $J_c$ are supported by the high pinning potentials of this compound. This slow change in $J_c$ as a function of magnetic field and temperature suggests that the $(Ba,K)Fe_2As_2$ superconductor has a superior $J_c$ field dependence, which is very beneficial for potential applications in high magnetic fields.

Partial flux jumps can be observed in magnetization loops if the superconducting sample is sufficiently large, the $J_c$ sufficiently high, the sample specific heat small and the ramp rate of the magnetic field sufficiently fast. Flux jumps were observed in our big $(Ba,K)Fe_2As_2$ single crystals, but at temperatures below 7 K. The shapes of these jumps are similar to the ones that have previously been observed and modelled in melt-textures YBCO samples[21]. So far, the occurrence of flux jumps has not been reported in any other Fe-based superconductors.

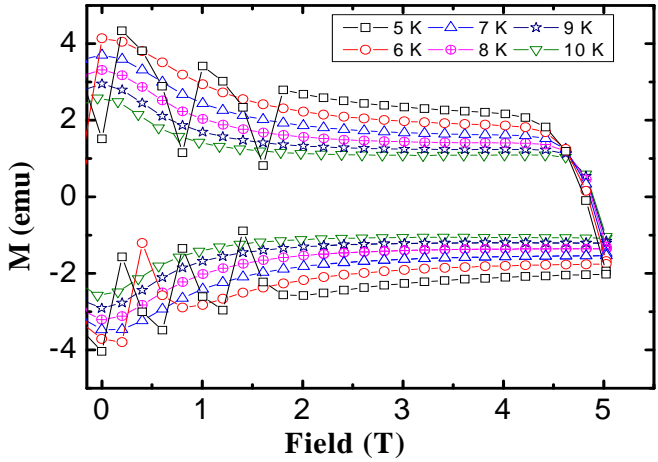

Fig.6. Magnetic hysteresis of a large $Ba_{0.72}K_{0.28}Fe_2As_2$ crystal.

Scanning electron microscopy (SEM) revealed that the 122 crystal has a typical 2D crystal structure (Inset (a) in Fig. 7). According to our TEM examination, we did not observe any noticeable crystal defects which could be possible pinning centers for the vortices. One of the typical atomic images taken along the (001) direction is presented in Fig. 7, which shows the typical features of square lattices on the (110) plane. All the surface atoms are perfectly ordered, and no superstructures are observed, as indicated by the selected large area electron diffraction pattern (Inset (b) in Fig. 7). These results reveal that the 122 crystal has a very high crystallinity and contains no anycrystal defects. Usually, the pinning is weak in a superconducting crystal with very high crystallinity. Now the question is where the strong pinning comes from for the perfect 122 crystal. There are two possible reasons. First, the K doping turns the parent compound from non-superconducting into superconducting and it introduces imperfections into the crystal lattice simultaneously, due to its different ionic size from the $Ba^{2+}$. Those lattice imperfection may act as intrinsic pinning centers in the $(Ba,K)Fe_2As_2$. That is to say, both superconductivity and the very strong intrinsic pinning are induced by K substitution. It has been reported that for the Fe-based superconductors, superconducting gaps vary with the characteristic length scales of the deviations from the average gap value.[22] These variations of the gap magnitude, the zero-bias conductance, and the coherence peak strength are quite well matched to each other. Moreover, these are further correlated with the average separation between the K dopant ions in the superconducting FeAs planes, which implies that the strength of the superconductivity is not uniform throughout the FeAs planes. The inhomogeneous superconducting strength may act as a source of pinning centers once magnetic flux exists inside the superconductor.

It should be pointed out that the nearly isotropic superconductivity[9] at low temperatures, which is also shown in the inset of Fig. 2, is the second reason responsible for the strong pinning in the $(Ba,K)Fe_2As_2$. It also gives the answer to the question why the $U_0$ value of $10^4$ K for $(Ba,K)Fe_2As_2$ is higher than that of $NdO_{0.82}F_{0.18}FeAs$, in which $U_0$ was 2000-3000 K[8]. It can be understood in terms of the higher superconducting coupling between the FeAs superconducting layers in FeAs-122 phase than in FeAs-1111 phase. In FeAs-122 phase, there are two superconducting FeAs layers, and the coupling between the superconducting layers is much stronger than in FeAs-1111 phase, which results in higher pinning potential and almost isotropic superconductivity as well as very rigid vortices in the FeAs-122 phase. That is to say, despite of the two dimensional nature of crystal structure of the 122 phase, the vortices can be easily pinned by any pinning centers regardless of directions. The isotropic rigid vortices observed in the two dimensional $(Ba,K)Fe_2As_2$ distinguish this compound from 2D high $T_c$ cuprate superconductors with 2D vortices, leading to this superconductor having intrinsic strong ability to carry large supercurrent in high magnetic fields.

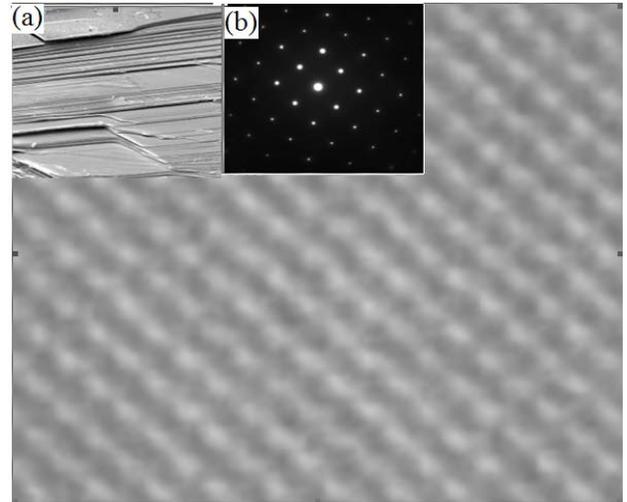

Fig. 7. High resolution TEM image of a $Ba_{0.72}K_{0.28}Fe_2As_2$ crystal. The insets contain a SEM image of the crystal surface (a) and an electron diffraction pattern along the (001) direction (b).

In conclusion, we found that the $(Ba,K)Fe_2As_2$ crystal shows very high intrinsic flux pinning strength, almost field independent high values of critical current density, high pinning potential of $10^4$ K, high $H_{c2}$, high $H_{irr}$, and low values of anisotropy of 1-3. The obtained $U_0$ values are record high compared to any existing superconducting single crystal. The isotropic rigid vortices observed in the two dimensional $(Ba,K)Fe_2As_2$ distinguish this compound from 2D high $T_c$ cuprate superconductors with 2D vortices. We observed the vortex avalanches in $(Ba,K)Fe_2As_2$ due to high $J_c$ which further supports the high value of the pinning potential *and the intrinsic strong flux pinning* in this compound. It is the K subsitition that induce both isotropic superconductivity and the very strong intrinsic pinning in the 122 compounds.

Acknowledgements
This work is supported by the Australian Research Council through ARC Discovery projects and by MIST/KRF (2009-0051705) of Korea.

*Email: xiaolin@uow.edu.au
** Email: silee77@sogang.ac.kr